\documentclass[sigconf]{acmart}
\usepackage{enumitem}
\usepackage{multirow}
\usepackage{todonotes}
\usepackage{color}
\usepackage{booktabs}

\newcommand\mycheck[1]{\textcolor{red}{#1}}
\setcopyright{acmcopyright}

\acmConference[JCDL '17]{The ACM/IEEE-CS Joint Conference on Digital Libraries}{June 2017}{Toronto, Ontario, Canada}
\copyrightyear{2016}
\acmPrice{15.00}
\acmISBN{123-4567-24-567/08/06}
\acmDOI{10.475/1234}

\begin{document}
\title{Citation sentence reuse behavior of scientists: A case study on\\ massive bibliographic text dataset of computer science}

\author{Mayank Singh}
\affiliation{%
  \institution{Dept.  of Computer Science and Engg.}
  \streetaddress{IIT Kharagpur, India}
  }
\email{mayank.singh@cse.iitkgp.ernet.in}

\author{Abhishek Niranjan}
\affiliation{%
  \institution{Dept.  of Computer Science and Engg.}
  \streetaddress{IIT Kharagpur, India}
  }
\email{aniranjan@cse.iitkgp.ernet.in}

\author{Divyansh Gupta}
\affiliation{%
  \institution{Dept.  of Computer Science and Engg.}
  \streetaddress{IIT Kharagpur, India}
  }
\email{divyansh.gupta@cse.iitkgp.ernet.in}

\author{Nikhil Angad Bakshi}
\affiliation{%
 \institution{Dept. of Mechanical Engg.}
  \streetaddress{IIT Kharagpur, India}
}
\email{nabakshi@iitkgp.ac.in}

\author{Animesh Mukherjee}
\affiliation{%
  \institution{Dept.  of Computer Science and Engg.}
  \streetaddress{IIT Kharagpur, India}}
  
\email{animeshm@cse.iitkgp.ernet.in}

\author{Pawan Goyal}
\affiliation{%
  \institution{Dept.  of Computer Science and Engg.}
  \streetaddress{IIT Kharagpur, India}
  }
\email{pawang@cse.iitkgp.ernet.in}

\renewcommand{\shortauthors}{Singh et al.}

\begin{abstract}
Our current knowledge of scholarly plagiarism is largely based on the similarity between full text research articles. In this paper, we propose an innovative and novel conceptualization of scholarly plagiarism in the form of reuse of explicit citation sentences in scientific research articles. Note that while full-text plagiarism is an indicator of a gross-level behavior, copying of citation sentences is a more nuanced micro-scale phenomenon observed even for well-known researchers. The current work poses several interesting questions and attempts to answer them by empirically investigating a large bibliographic text dataset from computer science containing millions of lines of citation sentences. In particular, we report evidences of massive copying behavior. We also present several striking real examples throughout the paper to showcase widespread adoption of this undesirable practice. In contrast to the popular perception, we find that copying tendency increases as an author matures. The copying behavior is reported to exist in all fields of computer science; however, the theoretical fields indicate more copying than the applied fields.
\vspace{-0.2cm}
\end{abstract}

\maketitle

\vspace{-0.5cm}
\section{Introduction}
\vspace{-0.1cm}
Scholarly text ``reuse" detection is a well known problem. 
It has received even more attention in the past decade due to overwhelming increase in the literature volume and ever increasing cases of plagiarism~\cite{judge2008plagiarism}. Traditionally, the focus has always been in the analysis of full text of the research articles. 
A highly celebrated work on copy detection by Brin et al.~\cite{brin1995copy} proposed a working system, called COPS, that detects copies (complete or partial) of research articles. They also proposed algorithms and metrics required for evaluating detection mechanisms. Bao et al.~\cite{bao2004semantic} applied Semantic Sequence Kin for document copy detection. Zu et al.~\cite{zu2006intrinsic} tried to detect plagiarism if references are not given. They develop a taxonomy of plagiarism delicts along with features for the quantification of style aspects.  
Recently, Citron et al.~\cite{citron2015patterns} described three classes of text reuse. They studied text reuse via a systematic pairwise comparison of the text content of all articles submitted to arXiv.org between 1991 -- 2012. They report that in some countries 15\% of submissions are detected as containing duplicated material. Lesk ~\cite{lesk2015many} presented scope of plagiarism within arXiv. They concluded that arXiv is now identifying the papers that have substantial overlap and is waiting to see if that affects the submission.

In recent past, numerous cases of widespread plagiarism have been detected leading to severe consequences. For example, Professor Matthew Whitaker from Arizona State University was made to resign after a series of plagiarism controversies\footnote{Wikipedia article: https://en.wikipedia.org/wiki/Matthew\_C.\_Whitaker\#Controversy}
. Professor Shahid Azam from University of Regina  has been accused for plagiarizing his own student master's thesis~\cite{bailey}. Leading to an even more disastrous consequence,  a city court in India briefly sent former vice-chancellor to jail on allegations of plagiarism~\cite{bailey}.

Plagiarism detection is considered computationally demanding as majority of proposed techniques rely on similarity between full text research articles.
In this paper, we propose a more nuanced micro-scale copying phenomenon by examining crowdsourced data generated in the form of explicit citation sentences. In contrast to full-text plagiarism that exhibits a gross-level behavior, we present a first attempt to understand the copying of citation sentences that corresponds to a more nuanced micro-scale phenomenon. 
We believe that researchers should be capable to describe previous literature without plagiarizing. To improve the quality and innovation, \vspace{-1cm} scientific community should strongly discourage such activities.

\vspace{-0.25cm}
\section{Dataset}
\label{sec:dataset}
In this paper, we use two computer science datasets crawled from Microsoft Academic Search (MAS)\footnote{http://academic.research.microsoft.com}. 
The first dataset~\cite{Chakraborty:2014} consists of bibliographic information (the title, the abstract, the keywords, its author(s), the year of publication, the publication venue, and the references) of more than 2.4 million papers published between 1859 -- 2012. The second dataset~\cite{Singh:2015:RCC:2806416.2806566} consists of more than 26 million citation sentences present all across the computer science articles published in the same time window. The scripts and processed data is available online\footnote{https://tinyurl.com/kzv4lhg} for download.

\vspace{-0.2cm}
\section{Citation sentences and similarity}
Throughout this paper we use the terms `citation sentence' and `citation context' interchangeably. If paper $P$ refers to paper $C$, then $P$ is termed the citing paper while $C$ is termed the cited paper. Given paper $P$, we consider those sentences as \textit{citation sentences} ($C_S$) that \textit{explicitly} cite the previous paper $C$. Note that, $P$ can refer to $C$ at many places in  the text leading to multiple $C_S$ for the same cited-citer pair. We process raw $C_S$ by replacing all citation placeholders (reference indexes, author names plus year etc.) with a single word ``CITATION''. We have been successful in replacing 16 different citation placeholder formats identified by Singh et al.~\cite{singh-EtAl:2016:COLING2}. A representative citation context from our dataset where~\cite{arapakis2009integrating} cites~\cite{han2005feature}, before and after pre-processing is as follows:

\noindent\textbf{Before:} \textit{Recommender systems are a personalized information filtering technology [4], designed to assist users in locating items of interest by providing useful recommendations.}

\noindent\textbf{After:} \textit{Recommender systems are a personalized information filtering technology CITATION, designed to assist users in locating items of interest by providing useful recommendations.}\\

\vspace{-0.15cm}
\noindent\textbf{Similarity computation}: This study massively relies on similarity scores between two citation contexts. Therefore, we utilize vector space model to compute similarity scores using tf-idf weighting scheme. We construct vocabulary from rich scientific text present in the second dataset. Due to computation complexity associated with similarity computations, out of  $\sim$1.5 million tokens, we only consider top 100,000 frequently occurring tokens. For each pair of $C_S$ vectors, similarity scores are generated using standard cosine similarity metric ($CosSim$). We employ python's machine learning library scikit-learn\footnote{http://scikit-learn.org/stable/} for all the computations.

\vspace{-0.1cm}
\section{Motivational statistics}
\label{sec:motivation}
We begin this work by examining the most intriguing and trivial question -- \textit{Whether} $C_S$ \textit{}{are really copied and to what extent?} To motivate the reader, we present a representative example of extensive copying from our dataset. Later in this section, we demonstrate that copying behavior is not a rare phenomenon; in contrast, it seems to have become a widespread convention.

\noindent\textbf{A representative example:} We found a large number of articles in our dataset whose incoming $C_S$ were partially or completely copied. In particular, we found cases where a paper receives an exact copy of $C_S$ from different citing papers. 
As a representative example, 
we found five copies of the citation context -- \textit{``We do not aim at formalizing some specific kind of state diagram, which has already successfully been done, see CITATION for example"}.
Here ``CITATION" placeholder consists of seven cited papers.

\begin{figure}[!tbh]
    \vspace{-0.3cm}
  \includegraphics[width=0.7\linewidth]{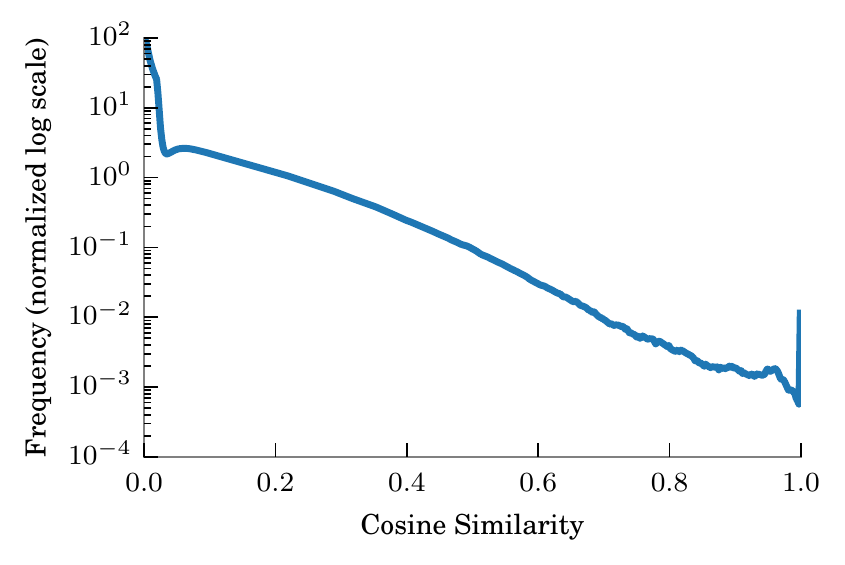}
    \vspace{-0.4cm}
  \caption{Cosine similarity score distribution for $C_S$ pairs.}
    \vspace{-0.3cm}
  \label{fig:CosineSimilarity}
\end{figure}

\noindent\textbf{Overall copying behavior:} Further, we attempted to understand the overall $C_S$ copying behavior. To start with, we randomly selected 24,800 papers. For each paper $p$, we compute pairwise cosine similarity between the incoming $C_S$. Figure~\ref{fig:CosineSimilarity} presents the distribution of similarity scores. As expected, most $C_S$ pairs have very low similarity scores ($CosSim\le0.2$). However, we also observe significant number of pairs having high similarity ($CosSim\ge0.8$). Most surprisingly, we found a sharp peak at $CosSim=1$; highlighting a very interesting observation that many $C_S$ are directly copied without any change. We also found papers where $C_S$ are copied from multiple articles. 
\\Overall, we found $\sim26$ thousand articles that consists of at least one citation context ditto copied ($CosSim = 1$) from another paper. We have used a strict metric for this; in specific, we concatenate the multiple instances for $C_S$ between the same pair of papers so that every pair has a unique citation context. Among this set, we find 148 articles each having at least five $C_S$, with $\ge60\%$ of $C_S$ being exact copy from previously published papers.

These initial statistics offer compelling evidence to study the phenomenon of copying $C_S$ in-depth. In the next section, we present extensive empirical analysis to understand the effect and properties of this interesting albeit undesirable phenomenon. 

 \vspace{-0.1cm}
\section{Large scale empirical study}
\label{sec:experiments}

In this section, we pose several interesting questions to better understand the characteristics of this micro-scale copying behavior. To answer these questions, we conduct in-depth empirical analysis on the computer science dataset described in section~\ref{sec:dataset}. Note that due to associated computation complexities in processing large text data, we perform individual experiments on smaller samples\footnote{We describe sample statistics in the beginning of each experiment.} of full dataset. For better interpretation, the posed questions are grouped into three categories: 1) Paperwise, 2) Authorwise, and 3) Fieldwise. We consider copying if pairwise cosine similarity between incoming $C_S$ is higher than 0.8.

\vspace{-0.2cm}
\subsection{Paperwise analysis}
\label{sec:paperexp}

\vspace{-0.1cm}
\subsubsection{Does publication age impact copying behavior?}
\label{sec:ageimpactscopy}
In this section, we attempt to investigate the temporal nature of the copying behavior. To start with, we select top 500 cited papers ($P$). For each selected paper $p \in P$, we study its incoming $C_S$ (that describe $p$) within $\Delta t$ years after publication. For the current study, we use $\Delta t = 1,2,3,4,5$.  Evidence suggests that copying behavior changes with the passage of time. More specifically, it first decreases and then starts increasing over the time-period with the exception for $\Delta t=1$. In each $y \in \Delta t$, we compute the fraction of $C_S$ copied (denoted by $f_q$) for each of the paper $q$, which cites paper $p \in P$. We compute the fraction of copied $C_S$ at $\Delta t$ year after publication by $F(\Delta t) = \frac{\sum_{q=1}^{n}f_q}{n}$. 
For each $\Delta t$, we compare similarity between $C_S$ generated in the year $\Delta t$ with all the $C_S$ generated between $\Delta (t=0)$ to $\Delta  (t-1)$. We find the value of average fraction of copied $C_S$ = 8.33\% in the first year after publication. In the next four successive years, the fraction varies as follows, 8.97\%, 8.02\%, 8.52\%, 9.12\%, indicating that the copying behavior decreases first and then gradually increases.

\vspace{-0.3cm}
\subsubsection{Are there differences in the cited versus the non-cited copying?}
\begin{figure}[!tbh]
  \vspace{-0.3cm}
  \includegraphics[width=1\linewidth]{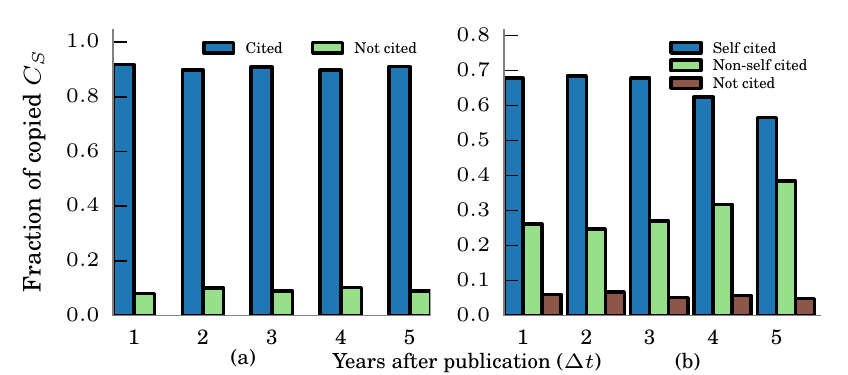}
  \vspace{-0.6cm}
  \caption{
a) Fraction of copied $C_S$ in two subsets, namely, cited and not cited. The fraction of two subsets remains nearly the same over the years. 
b) Fraction of $C_S$ in three subsets, namely, self cited, non-self cited and not cited. 
The fraction of \textit{self cited} subset decline over the years leading to increase in the fraction of \textit{non-self cited} subset.
}
  \vspace{-0.3cm}
   \label{fig:paperwise}
\end{figure}

Next, we divide copied $C_S$ into two subsets, namely, i) cited (\textit{CC}), and ii) not cited (\textit{NC}). \textit{CC} consists of copied $C_S$ where the source paper (from which context is being copied) is cited, whereas \textit{NC} consists of copied $C_S$ where the source paper is not cited. Figure~\ref{fig:paperwise}a reports proportion of copied $C_S$ into these subsets at five time periods after publication. To our surprise, we find that fraction of two subsets remains same over the years. Fraction of \textit{NC} is significantly lower than fraction of \textit{CC}. 

The copy and cited behavior is a combination of two distinct forms of copying, namely, \textit{self copying} and \textit{non-self copying}. In self copying, an author copies her own older $C_S$, whereas in non-self copying, an author copies $C_S$ written by other authors. Therefore, we split \textit{CC} subset into further two subsets, namely, i) self cited (\textit{SC}), and ii) non-self cited (\textit{NSC}). Figure~\ref{fig:paperwise}b reports proportion of copied $C_S$ into three subsets ($SC$, $NSC$ and $NC$) at five time periods after publication. The single most striking observation from Figure~\ref{fig:paperwise}b is that fraction of \textit{SC} is much higher than \textit{NSC} indicating that majority of the $C_S$ are copied by an author in their own future publications. However, as authors' interest shifts from one topic to other, the fraction of \textit{SC} declines resulting into increase in the fraction of \textit{NSC}.

\vspace{-0.2cm}
\subsection{Authorwise analysis}
\label{sec:authexp}

\subsubsection{Does an author's increasing experience influence her copying behavior?}
In this section, we present empirical results to prove that researchers in their early stage of academic career behave differently than after gaining experience. We begin this analysis by selecting 7175 random authors that have at least 20 years of citation history. For each author, we compute fraction of copied $C_S$ from previously published papers. Figure~\ref{fig:Authorwise}a presents average fraction of copied $C_S$ over 20 years of the author life span. In contrast to the popular perception, we observe that the copying tendency increases as an author matures.

\subsubsection{Does an author's popularity influence his copying behavior?}
We observe that an author's popularity plays a critical role in influencing his copying patterns. For this study, we select authors with varying popularity but with similar academic age\footnote{In order to keep the author experience same.}. We select all authors that started their career from the year 1995. We compute the authors' h-index (in 2012) to measure their individual popularity. To better visualize the influencing behavior, we divide the authors into three h-index buckets:
\begin{itemize}[noitemsep,nolistsep]
 \item \textbf{Bucket 1:} h-index $<5$
 \item \textbf{Bucket 2:} $5\leq$ h-index $<15$
 \item \textbf{Bucket 3:} h-index $\geq15$
\end{itemize}

\begin{figure}[!tbh]
  \includegraphics[width=1\linewidth]{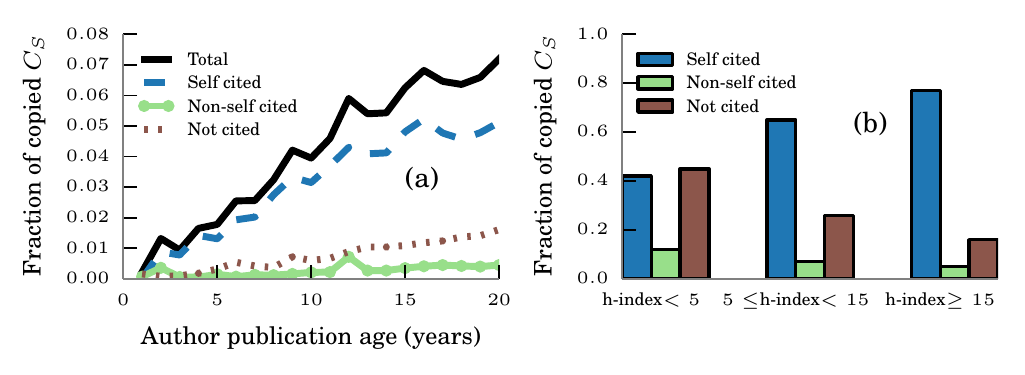}
  \vspace{-0.85cm}
  \caption{
  a) Average fraction of copied $C_S$ over 20 years of author life span. Copying tendency increases as author matures.
  b) Proportion of three copying behaviors in three h-index buckets. Popular authors prefer to copy their own $C_S$.
  }
   \vspace{-0.3cm}
  \label{fig:Authorwise}
\end{figure}

Here \textbf{Bucket 1} represents the least popular authors whereas \textbf{Bucket 3} consists of the most popular authors. We compute the fraction of copied $C_S$ for each author in each bucket and present aggregated statistics. We observe that the most popular authors (\textbf{Bucket 3}) show maximum copying tendency. Whereas least popular authors (\textbf{Bucket 1}) show least copying tendency. On average, 2.07\% $C_S$ of \textbf{Bucket 1} are copied from previous papers. This fraction increases to 4.17\% and 6.16\% for \textbf{Bucket 2} and \textbf{Bucket 3} respectively. To investigate in more detail, we divide copied $C_S$ into three subsets as described in section~\ref{sec:ageimpactscopy}. Figure~\ref{fig:Authorwise}b presents proportion for three different copying behaviors in three h-index buckets. Fraction of \textit{SC} increases as h-index increases. Popular authors prefer to copy their own $C_S$ while less popular authors try to copy $C_S$ written by other authors. Figure~\ref{fig:Authorwise}b reports significant proportion of \textit{NC} in \textbf{Bucket 1}. 
We observe similar results if we consider the author's increasing publication count (in place of h-index) on her copying behavior.

\vspace{-0.3cm}

\vspace{-0.35cm}
\subsection{Fieldwise Analysis}
\label{sec:fieldexp}
As described in section~\ref{sec:dataset}, our dataset consists of papers from 24 fields of computer science. In this section, we attempt to demonstrate the copying behavior in different fields of research. We randomly select 20,000 research papers from 24 distinct fields of computer science. The distribution of $C_S$ in each field is shown in Figure~\ref{fig:FieldWiseCos}a. Two interesting questions that require fieldwise analysis are:

\begin{figure*}[!tbh]
  \vspace{-0.52cm}
  \includegraphics[width=1\linewidth]{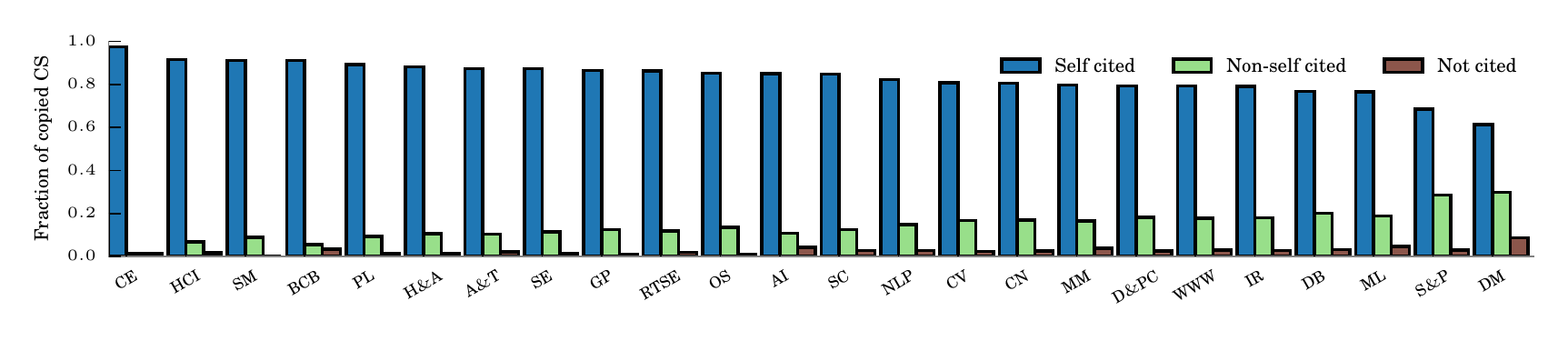}
  \vspace{-1.1cm}
  \caption{Proportions of three copying behaviors for 24 computer science fields. Majority of the copied contexts are originating from self citations. Applied fields have more tendency of copying and not citing than theoretical fields.}
  \label{fig:FieldWiseauth}
\vspace{-0.4cm}
\end{figure*}

\subsubsection{Is paper-wise copying behavior different in different computer science fields?}
Similar to the experiment in section~\ref{sec:ageimpactscopy}, for each field, we divide copied $C_S$ into three categories, \textit{SC}, \textit{NSC} and \textit{NC}. Figure~\ref{fig:FieldWiseauth} presents proportion of three categories for 24 computer science fields. 
For all fields, majority of the copied contexts are originating from self citations. Next major proportion goes to non-self citations.
Overall, we observe that applied fields have more tendency of copying and not citing than theoretical fields. 

\vspace{-0.1cm}
\subsubsection{{Is copying behavior same across all computer science fields?}}
\begin{figure}[!tbh]
  \vspace{-0.4cm}
  \includegraphics[width=1\linewidth]{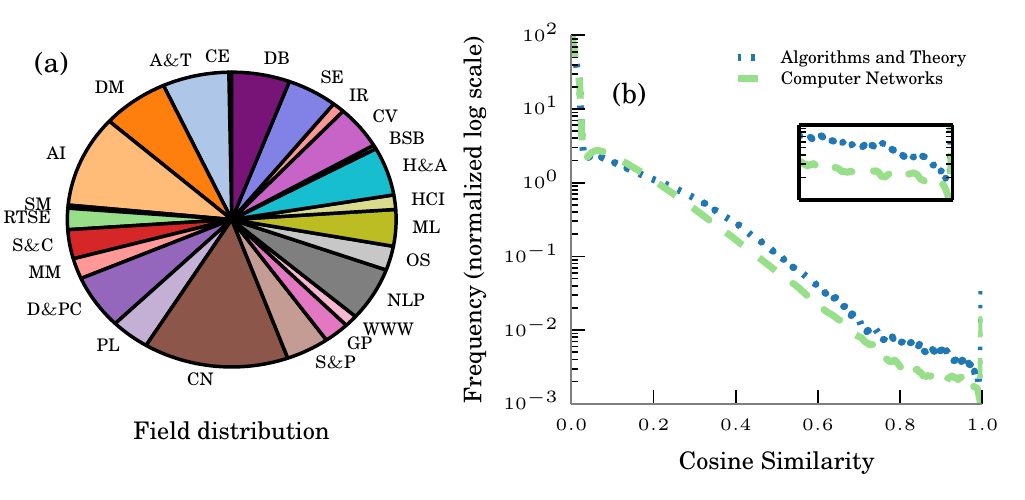}
  \vspace{-0.9cm}
  \caption{
  Fieldwise analysis: 
  a) Distribution of $C_S$ sampled from 24 fields. b) Comparison between cosine similarity distribution of two representative computer science fields. Inset shows significant difference between Algorithms \& Theory (theoretical) and Computer Networks (applied) for higher $CosSim (\geq0.8)$ values.}
  \label{fig:FieldWiseCos}
  \vspace{-0.5cm}
\end{figure}

We perform this experiment along similar lines as the motivational study (see section~\ref{sec:motivation}), except that now papers are divided into 24 computer science fields\footnote{http://tinyurl.com/n2rwkbs}. Figure~\ref{fig:FieldWiseCos}b presents cosine similarity distribution for two representative fields. We were surprised to find clear demarcation between theoretical fields\footnote{https://en.wikipedia.org/wiki/Computer\_science} like, Algorithms \& Theory etc., and applied fields like, Computer Networks etc. Even though, for small values of $CosSim$ , all fields show similar behavior, for higher $CosSim (\geq0.8)$ values (see inset Figure~\ref{fig:FieldWiseCos}), theoretical fields show higher copying behavior as compared to applied fields. Note that, a sharp peak at $CosSim=1$ is observed across all fields. 

\vspace{-0.1cm}
\section{Conclusions}
This paper has investigated micro-scale phenomenon of text reuse in scientific articles.  
Throughout the paper,  we pose several interesting research questions and present an in-depth empirical analysis to answer them. We have provided further evidence that the theoritical fields indicate more copying than the applied fields. 
Finally, a number of potential limitations need to be considered. First, the current study employs a computer science dataset only. In the future, we plan to extend this study to other research fields as well. Second, cosine similarity metric may be too simplistic for this complex study. To further our research, we plan to employ word embeddings for more meaningful similarity computations. 
On account of the fact that the current work is only a preliminary attempt to understand micro-scale copying phenomenon of citation sentences, future extensions could possibly lead to rescaling of several popularity metrics such as h-index, impact factor etc. as number of times a paper has been cited may not be a true metric for impact of a paper or researchers in the research community. Alternatively, if this behavior is kept in check the quality of research can be expected to improve and the current popularity metrics will conform to the intuition behind their meaning.

\vspace{-0.2cm}
\bibliographystyle{ACM-Reference-Format}
\bibliography{sigproc}

\end{document}